\documentclass[11pt,a4paper]{article}

\usepackage[T1]{fontenc}
\usepackage[utf8]{inputenc}
\usepackage{lmodern}
\usepackage[a4paper,margin=1in]{geometry}
\usepackage{microtype}
\usepackage{amsmath,amssymb,amsfonts,mathtools,mathrsfs}
\usepackage{amsthm}
\usepackage{bm}
\usepackage{graphicx}
\usepackage{booktabs}
\usepackage{enumitem}
\usepackage{caption}
\usepackage{float}
\usepackage{url}
\usepackage[hidelinks]{hyperref}
\usepackage[nameinlink,capitalise,noabbrev]{cleveref}

\graphicspath{{figures/}}
\allowdisplaybreaks
\newtheorem{theorem}{Theorem}[section]
\newtheorem{proposition}[theorem]{Proposition}
\newtheorem{lemma}[theorem]{Lemma}
\newtheorem{corollary}[theorem]{Corollary}
\theoremstyle{remark}
\newtheorem{remark}[theorem]{Remark}

\crefname{equation}{equation}{equations}
\Crefname{equation}{Equation}{Equations}
\crefname{proposition}{proposition}{propositions}
\Crefname{proposition}{Proposition}{Propositions}
\crefname{corollary}{corollary}{corollaries}
\Crefname{corollary}{Corollary}{Corollaries}
\crefname{lemma}{lemma}{lemmas}
\Crefname{lemma}{Lemma}{Lemmas}
\crefname{theorem}{theorem}{theorems}
\Crefname{theorem}{Theorem}{Theorems}
\crefname{figure}{figure}{figures}
\Crefname{figure}{Figure}{Figures}
\crefname{table}{table}{tables}
\Crefname{table}{Table}{Tables}
\crefname{section}{section}{sections}
\Crefname{section}{Section}{Sections}

\newcommand{\R}{\mathbb R}
\newcommand{\M}{\mathcal M}
\newcommand{\EE}{\mathbb E}
\newcommand{\PP}{\mathbb P}
\newcommand{\Spp}[1]{\mathbb S_{++}^{#1}}
\newcommand{\tr}{\operatorname{tr}}
\newcommand{\Log}{\operatorname{Log}}

\newcommand{\argmin}{\operatorname*{arg\,min}}
\DeclareMathOperator{\Cov}{Cov}
\DeclareMathOperator{\Var}{Var}

\title{Poisson-Sampled Fr\'echet Means on Gaussian Information Manifolds}
\author{Gourab Ghatak\\
\small Department of Electrical Engineering, Indian Institute of Technology Delhi\\
\small New Delhi 110016, India\\
\small \texttt{gghatak@ee.iitd.ac.in}}
\date{}

\begin{document}
\maketitle

\begin{abstract}
Distribution-valued marks in a spatial network live on a statistical manifold, whereas their sampling locations are governed by stochastic geometry.  We develop a rigorous finite-window theory for Fr\'echet means sampled by a Poisson point process.  The population target is the barycenter of the normalized window mark law, avoiding the typically divergent unnormalized objective on the whole space.  Conditional on the Poisson count, independently marked points form an ordinary independent sample; this yields an exact zero-truncated Poissonization transform for fixed-sample risks, tails, consistency, and intrinsic central limit theorems.  For geodesic-supported marks, we obtain closed-form random-count mean-square errors.  For a common spatial random field, the error separates exactly into a correlation floor and a Poisson term, and stationary Gaussian fields satisfy a spatial central limit theorem with variance equal to the integrated field covariance plus a Poisson diagonal contribution.  Slivnyak's theorem and Poisson splitting then give valid reduced- and non-reduced Palm laws and exact resampling and thinning formulas; the correlation floor cancels when the retained and full barycenters share the same field realization.  The theory is specialized to covariance-varying Gaussian models: the full univariate Fisher--Rao manifold and the affine-invariant manifold of multivariate covariance matrices, including arbitrary noncommuting, spatially heterogeneous covariance laws.  The Wasserstein discussion records the correct one-dimensional covariance barycenter and isolates its commuting flat boundary case.
\end{abstract}

\medskip

\section{Introduction}
\label{sec:introduction}

A spatial network may carry substantially richer marks than scalar powers, loads, or fading coefficients.  A sensor, agent, or local inference engine can instead attach a probability distribution to its location, for example a Gaussian posterior
\(p_x=\mathcal N(\mu_x,\Sigma_x)\).  The spatial configuration is naturally modeled by a point process, while the marks live on a statistical manifold.  Averaging such marks is therefore simultaneously a problem in stochastic geometry and information geometry.

A direct combination of the two theories is deceptively delicate.  Three issues arise before any geometric calculation is attempted.  First, a stationary Poisson point process on all of \(\R^s\) has infinitely many points; the unnormalized integral of squared mark-to-candidate distances is consequently infinite in most nondegenerate models.  Second, a bounded observation window contains a Poisson--distributed number of marks and is empty with positive probability.  Third, Campbell's formula identifies the expectation of a random sum for each fixed candidate point, but it does not by itself establish convergence of random minimizers or justify interchanging minimization and expectation.  These points require a finite-window population law, explicit treatment of the zero-count event, and a random-index argument based on fixed-sample Fr\'echet-mean theory.

The geometry also matters.  Restricting Gaussian marks to a common covariance matrix makes the induced mean-only geometry flat, and hence removes most of the information-geometric content.  More seriously, the common-covariance set is not geodesically closed in the full Fisher--Rao manifold: even two univariate Gaussians with the same variance have a full-manifold geodesic midpoint with a larger variance.  Nontrivial covariance variation must therefore be retained if the interaction with information geometry is to be genuine.

\paragraph{Relation to existing theory.}

Fr\'echet means on complete nonpositively curved spaces are well established: finite second moments give a unique barycenter, empirical barycenters satisfy laws of large numbers, and smooth Riemannian models admit intrinsic central limit theorems \cite{sturm2003,bhattacharya2003,bhattacharya2005,pennec2006}.  Those results are formulated for a deterministic sample size.  The present paper does not replace that theory; it gives an exact interface through which any valid fixed-sample result can be transferred to a zero-truncated Poisson sample and then separates this random-count effect from spatial heterogeneity and spatial dependence.

On the stochastic-geometry side, the conditional distribution of Poisson points in a bounded window, Campbell--Mecke identities, Slivnyak's theorem, and independent thinning are classical \cite{kingman1993,lastpenrose2017,chiu2013}.  Their role here is more specific than first-moment averaging.  Conditioning on the count identifies the correct mark law, second-order integration exposes the correlation floor of a common random field, Slivnyak distinguishes reduced from non-reduced Palm estimators, and Poisson splitting preserves the dependence between a thinned sample and its parent sample.

The Fisher geometry of the full non-centered multivariate Gaussian family has been studied through its Riemannian structure, embeddings, geodesic equations, special cases, bounds, and numerical approximations \cite{skovgaard1984,calvo1990,calvo1991,kobayashi2023,nielsen2023,pinele2020,miyamoto2024}.  A simple global closed form for arbitrary endpoint pairs is nevertheless unavailable.  We therefore work with two Gaussian information manifolds on which the required geometry is exact: the hyperbolic manifold of all univariate Gaussians and the affine-invariant manifold of multivariate covariances at fixed mean \cite{amari2000,costa2015,bhatia2007,moakher2005}.  Gaussian Wasserstein barycenters obey a different covariance geometry and a nonlinear fixed-point equation \cite{takatsu2011,malago2018,alvarez2016}.  We keep these geometries separate and use each only where the curvature and geodesic assumptions required by the proofs are valid.

This paper develops a focused theory of \emph{Poisson-sampled Fr\'echet means} on nonpositively curved information manifolds.  The main results are as follows.

\begin{enumerate}[label=\textbf{C\arabic*.},leftmargin=2.2em]
\item We formulate the population barycenter in a bounded window through the normalized mixture law of the marks.  Conditional on the Poisson count, independently marked points become an ordinary independent and identically distributed sample from this law.  This yields well-posedness, consistency under intensity scaling, and a correct condition for expanding-window limits.

\item We introduce an exact Poissonization operator that converts any fixed-sample risk or tail bound into its random-count counterpart.  The reciprocal-count factor is
\[
H(m):=\EE\!\left[\frac{1}{N}\,\middle|\,N>0\right]
=\frac{\operatorname{Ei}(m)-\gamma-\log m}{e^m-1},
\qquad N\sim\operatorname{Poisson}(m),
\]
not \(1/m\).  We also establish a random-index transfer of intrinsic central limit theorems.

\item For marks supported on a geodesic, the manifold problem becomes exactly one-dimensional in arclength coordinates.  This gives closed-form mean-square errors and concentration bounds.  For a spatially correlated geodesic field, the error separates into a correlation floor and a Poisson sampling term.  Increasing point density cannot remove the correlation floor on a fixed window, whereas expanding the window produces a different asymptotic constant involving the integrated covariance of the field.  For stationary Gaussian fields, we further prove an intrinsic spatial central limit theorem with this same variance constant.

\item We derive valid reduced- and non-reduced Palm statements using Slivnyak's theorem.  The non-reduced Palm estimator, which includes the typical point, has an exact risk involving two Poisson count factors.  We also derive exact laws for independent resampling and Poisson thinning of a spatially correlated field.  The field-level correlation floor cancels when the retained and full barycenters share the same realization, a dependence that is absent from two independently generated fields.

\item We specialize the theory to two genuinely curved Gaussian models: the full univariate Fisher--Rao manifold, which is a scaled hyperbolic half-plane, and the fixed-mean multivariate covariance manifold endowed with the affine-invariant Fisher metric.  The latter accommodates arbitrary, including noncommuting, covariance matrices.  A short Wasserstein discussion identifies the commuting flat boundary case and records the correct one-dimensional covariance barycenter.
\end{enumerate}

The paper intentionally does not attach a universal ``semantic'' meaning to barycenter preservation, nor does it propose a bandit policy or compare a geometric estimator with a baseline that discards covariance information.  Such applications require their own task-specific losses and statistically matched baselines.  Here the aim is the mathematical interface itself: random spatial sampling, intrinsic averaging, correlation, Palm conditioning, and thinning on curved Gaussian manifolds.

\paragraph{Organization.}
\Cref{sec:preliminaries} gives the finite-window model and its well-posed population target.  \Cref{sec:poissonization} develops the Poissonization calculus.  Exact geodesic and spatial-correlation laws are established in \cref{sec:geodesic,sec:correlated}.  Palm and thinning results appear in \cref{sec:palm-thinning}.  Gaussian information-geometric specializations are given in \cref{sec:gaussian}.  \Cref{sec:numerics} validates the formulas numerically, and \cref{sec:discussion} discusses scope and limitations.

\section{Hadamard barycenters and the marked Poisson model}
\label{sec:preliminaries}

\subsection{Fr\'echet means in a Hadamard space}

Let \((\M,d)\) be a complete separable Hadamard space, equivalently a complete CAT(0) space.  Thus every two points \(p,q\in\M\) are joined by a unique constant-speed geodesic \(\gamma_{p,q}:[0,1]\to\M\), and
\begin{equation}
\label{eq:cat0}
 d^2\!\left(z,\gamma_{p,q}(t)\right)
 \le (1-t)d^2(z,p)+t d^2(z,q)-t(1-t)d^2(p,q)
\end{equation}
for all \(z\in\M\) and \(t\in[0,1]\); see \cite{bridson1999,sturm2003}.  The nonpositive curvature in \eqref{eq:cat0} implies global geodesic convexity of squared distance.

For a Borel probability measure \(Q\) on \(\M\) with finite second moment, define
\begin{equation}
\label{eq:frechet-functional}
 F_Q(q):=\int_{\M} d^2(p,q)\,Q(\mathrm dp).
\end{equation}
Its unique minimizer
\begin{equation}
\label{eq:barycenter-def}
 b(Q):=\argmin_{q\in\M}F_Q(q)
\end{equation}
is the Fr\'echet mean or barycenter of \(Q\) \cite{frechet1948,sturm2003}.  Two standard facts will be used repeatedly.  First, the variance inequality
\begin{equation}
\label{eq:variance-ineq}
 F_Q(q)-F_Q\bigl(b(Q)\bigr)\ge d^2\!\left(q,b(Q)\right)
\end{equation}
holds for every \(q\in\M\).  Second, the barycenter map is contractive: for probability measures \(Q_1,Q_2\) with finite second moments,
\begin{equation}
\label{eq:barycenter-contraction}
 d\!\left(b(Q_1),b(Q_2)\right)\le W_1(Q_1,Q_2)\le W_2(Q_1,Q_2),
\end{equation}
where the Wasserstein distances in \eqref{eq:barycenter-contraction} are built from the metric \(d\) on \(\M\) \cite{sturm2003}.  Hence a population barycenter is stable under convergence of its mark law, without any minimization--expectation interchange.

For deterministic observations \(P_1,\ldots,P_n\in\M\), \(n\ge1\), write
\begin{equation}
\label{eq:empirical-barycenter-fixed-n}
 \widehat b_n:=b\!\left(\frac1n\sum_{i=1}^n\delta_{P_i}\right)
 =\argmin_{q\in\M}\frac1n\sum_{i=1}^n d^2(P_i,q).
\end{equation}
The normalizing factor \(1/n\) does not change the minimizer, but it is useful when comparing empirical and population functionals.  Laws of large numbers and central limit theorems for intrinsic means are classical in suitable metric and Riemannian settings \cite{sturm2003,bhattacharya2003,bhattacharya2005,pennec2006}.

\subsection{A finite-window distribution-marked Poisson process}

Let \(B\subset\R^s\) be a bounded Borel set and let \(\Lambda\) be a locally finite diffuse intensity measure satisfying
\begin{equation}
\label{eq:m-def}
 0<m_B:=\Lambda(B)<\infty.
\end{equation}
Let \(\Phi\) be a Poisson point process on \(\R^s\) with intensity measure \(\Lambda\), and write \(N_B:=|\Phi\cap B|\).  At a point \(x\in\Phi\), attach an \(\M\)-valued mark \(P_x\).  In the independently marked model, conditional on the locations, the marks are independent with location-dependent kernel \(Q_x(\mathrm dp)\).  Assume that \(x\mapsto Q_x(A)\) is measurable for every Borel \(A\subset\M\) and that, for some \(o\in\M\),
\begin{equation}
\label{eq:second-moment-kernel}
 \frac1{m_B}\int_B\int_{\M}d^2(p,o)\,Q_x(\mathrm dp)\,\Lambda(\mathrm dx)<\infty.
\end{equation}
Define the normalized location law and the window-averaged mark law by
\begin{equation}
\label{eq:w-and-QB}
 w_B(\mathrm dx):=\frac{\mathbf1_B(x)\Lambda(\mathrm dx)}{m_B},
 \qquad
 Q_B(\mathrm dp):=\int_B Q_x(\mathrm dp)\,w_B(\mathrm dx).
\end{equation}
The correct population target in the window is
\begin{equation}
\label{eq:window-population-barycenter}
 b_B:=b(Q_B)
 =\argmin_{q\in\M}
 \frac1{m_B}\int_B\int_{\M}d^2(p,q)Q_x(\mathrm dp)\Lambda(\mathrm dx).
\end{equation}
This is a probability-normalized objective on a finite window.  No integral over the entire plane is required.

On the event \(N_B>0\), define the empirical window barycenter
\begin{equation}
\label{eq:window-empirical-barycenter}
 \widehat b_B
 :=\argmin_{q\in\M}\frac1{N_B}\sum_{x\in\Phi\cap B}d^2(P_x,q).
\end{equation}
We leave \(\widehat b_B\) undefined on \(\{N_B=0\}\) and state all finite-window risks conditionally on \(N_B>0\).  This is preferable to assigning an arbitrary default point whose contribution would depend on the coordinate convention.

\begin{proposition}[Conditional sampling representation]
\label{prop:conditional-iid}
Under \eqref{eq:m-def}--\eqref{eq:second-moment-kernel}:
\begin{enumerate}[label=(\roman*)]
\item \(Q_B\) has a finite second moment and \(b_B\) in \eqref{eq:window-population-barycenter} exists uniquely.
\item Conditional on \(N_B=n\ge1\), the location--mark pairs \((X_i,P_i)_{i=1}^n\) in \(B\) are independent and identically distributed with joint law
\[
 w_B(\mathrm dx)Q_x(\mathrm dp).
\]
In particular, \(P_1,\ldots,P_n\) are independent and identically distributed with marginal law \(Q_B\).
\item Conditional on \(N_B=n\), \(\widehat b_B\) has exactly the same law as the fixed-sample barycenter \(\widehat b_n\) of \(n\) independent draws from \(Q_B\).
\end{enumerate}
\end{proposition}

\begin{proof}
The moment assertion follows by integrating \eqref{eq:second-moment-kernel}, and uniqueness follows from the Hadamard property.  The conditional distribution of Poisson points in a finite window is the normalized intensity law \(w_B\); the marking theorem then gives the stated joint law \cite{kingman1993,lastpenrose2017,chiu2013}.  Marginalizing over \(x\) yields \(Q_B\), and the final assertion follows from \eqref{eq:empirical-barycenter-fixed-n}.
\end{proof}

\begin{corollary}[Intensity-scaling consistency]
\label{cor:intensity-consistency}
Let \(\Phi_t\) have intensity measure \(t\Lambda\), with the mark kernel \(Q_x\) unchanged, and let \(\widehat b_{B,t}\) be defined as in \eqref{eq:window-empirical-barycenter}.  Then
\[
 \widehat b_{B,t}\longrightarrow b_B
 \quad\text{in probability conditional on }N_{B,t}>0,
 \qquad t\to\infty.
\]
\end{corollary}

\begin{proof}
Conditional on \(N_{B,t}=n\), Proposition~\ref{prop:conditional-iid} reduces the problem to the fixed-sample law of large numbers for empirical barycenters in a Hadamard space \cite{sturm2003}.  Since \(N_{B,t}\sim\operatorname{Poisson}(tm_B)\) tends to infinity in probability and \(\PP(N_{B,t}=0)=e^{-tm_B}\to0\), a random-index argument completes the proof.
\end{proof}

The population law may change when the window changes.  The following statement gives a sufficient condition for a genuine spatial limit.

\begin{proposition}[Expanding-window population limit]
\label{prop:window-limit}
Let \((B_R)\) be bounded windows with \(m_R:=\Lambda(B_R)\to\infty\), and let \(Q_{B_R}\) be defined by \eqref{eq:w-and-QB}.  Suppose that, for some probability measure \(Q_\infty\) with finite second moment,
\begin{equation}
\label{eq:QB-W2-limit}
 W_2(Q_{B_R},Q_\infty)\longrightarrow0.
\end{equation}
Then \(b_{B_R}\to b_\infty:=b(Q_\infty)\).  Moreover, if the process is independently marked and \(\widehat b_{B_R}\) is conditioned on \(N_{B_R}>0\), then
\[
 \widehat b_{B_R}\longrightarrow b_\infty
 \quad\text{in probability}.
\]
\end{proposition}

\begin{proof}
The first claim follows immediately from \eqref{eq:barycenter-contraction}.  It remains to control the empirical term although the sampling law changes with \(R\).  For each \(R\), choose an optimal \(W_2\)-coupling \((P_{R,i},P_{\infty,i})_{i\ge1}\), independently across \(i\), with marginals \(Q_{B_R}\) and \(Q_\infty\).  For every deterministic \(n\ge1\), barycenter contractivity gives
\[
 d\!\left(
 b\!\left(\frac1n\sum_{i=1}^n\delta_{P_{R,i}}\right),
 b\!\left(\frac1n\sum_{i=1}^n\delta_{P_{\infty,i}}\right)
 \right)
 \le
 \left(\frac1n\sum_{i=1}^n d^2(P_{R,i},P_{\infty,i})\right)^{1/2}.
\]
The squared expectation of the right-hand side is \(W_2^2(Q_{B_R},Q_\infty)\to0\).  The empirical barycenter of the \(P_{\infty,i}\)'s converges in probability to \(b_\infty\) as \(n\to\infty\).  Consequently, the triangular sequence with any deterministic \(n_R\to\infty\) converges to \(b_\infty\).  Finally \(N_{B_R}\to\infty\) in probability, and conditioning on \(N_{B_R}>0\) is asymptotically immaterial because \(e^{-m_R}\to0\).  A random-index mixture argument completes the proof.
\end{proof}

\begin{remark}
\label{rem:no-global-integral}
Proposition~\ref{prop:window-limit} separates two logically different limits: the sampling error \(d(\widehat b_{B_R},b_{B_R})\) and the population drift \(d(b_{B_R},b_\infty)\).  Campbell's theorem is useful for evaluating the normalized functional in \eqref{eq:window-population-barycenter}, but it neither supplies \eqref{eq:QB-W2-limit} nor proves convergence of minimizers.
\end{remark}

\section{Poissonization of fixed-sample Fr\'echet-mean theory}
\label{sec:poissonization}

Proposition~\ref{prop:conditional-iid} shows that the point process enters the independent-mark problem through a random sample size.  We now make this dependence exact.

For a nonnegative sequence \(a=(a_n)_{n\ge1}\), define the zero-truncated Poissonization operator
\begin{equation}
\label{eq:poissonization-operator}
 \mathscr P_m[a]
 :=\frac{1}{e^m-1}\sum_{n=1}^{\infty}\frac{m^n}{n!}a_n,
 \qquad m>0.
\end{equation}
If \(N\sim\operatorname{Poisson}(m)\), then \(\mathscr P_m[a]=\EE[a_N\mid N>0]\).

\subsection{Exact count factors}

Set
\begin{equation}
\label{eq:A-H-def}
 A(m):=\sum_{n=1}^{\infty}\frac{m^n}{n\,n!}
 =\operatorname{Ei}(m)-\gamma-\log m,
 \qquad
 H(m):=\frac{A(m)}{e^m-1}.
\end{equation}
Here \(\gamma\) is the Euler--Mascheroni constant and \(\operatorname{Ei}\) is the exponential integral.  Direct summation gives
\begin{equation}
\label{eq:H-expectation}
 H(m)=\EE\!\left[\frac1N\,\middle|\,N>0\right].
\end{equation}
Two additional factors needed for Palm sampling are
\begin{equation}
\label{eq:G1-G2}
 G_1(m):=\EE\!\left[\frac1{N+1}\right]=\frac{1-e^{-m}}{m},
 \qquad
 G_2(m):=\EE\!\left[\frac1{(N+1)^2}\right]=\frac{e^{-m}A(m)}{m}.
\end{equation}
The large-\(m\) expansion of \(H\) is
\begin{equation}
\label{eq:H-asymptotic}
 H(m)=\frac1m+\frac1{m^2}+\frac2{m^3}+O(m^{-4}).
\end{equation}
Thus \(1/m\) is only a first-order approximation.  It is not a valid finite-\(m\) upper bound; for example, \(H(5)\approx0.25777>0.2\).

\subsection{Risk, tails, and a random-index central limit theorem}

Let \(P_1,P_2,\ldots\) be independent and identically distributed with law \(Q\in\mathcal P_2(\M)\), let \(b=b(Q)\), and let \(\widehat b_n\) be as in \eqref{eq:empirical-barycenter-fixed-n}.  Independently, let \(N\sim\operatorname{Poisson}(m)\), and define \(\widehat b_N\) on \(\{N>0\}\).

\begin{theorem}[Exact Poissonization]
\label{thm:poissonization}
For every nonnegative measurable loss \(L:\M\times\M\to[0,\infty]\),
\begin{equation}
\label{eq:loss-poissonization}
 \EE\!\left[L(\widehat b_N,b)\,\middle|\,N>0\right]
 =\mathscr P_m\!\left[
 n\mapsto\EE L(\widehat b_n,b)
 \right].
\end{equation}
In particular, if
\[
 r_n:=\EE d^2(\widehat b_n,b),
 \qquad
 \alpha_n(u):=\PP\{d(\widehat b_n,b)\ge u\},
\]
then
\begin{align}
\label{eq:risk-poissonization}
 \EE\!\left[d^2(\widehat b_N,b)\,\middle|\,N>0\right]
 &=\mathscr P_m[(r_n)],\\
\label{eq:tail-poissonization}
 \PP\{d(\widehat b_N,b)\ge u\mid N>0\}
 &=\mathscr P_m[(\alpha_n(u))].
\end{align}
Consequently:
\begin{enumerate}[label=(\roman*)]
\item If \(r_n\le C/n\), then
\begin{equation}
\label{eq:poissonized-risk-bound}
 \EE[d^2(\widehat b_N,b)\mid N>0]\le C H(m).
\end{equation}
\item If \(\alpha_n(u)\le C_0e^{-n\psi(u)}\), where \(\psi(u)\ge0\), then
\begin{equation}
\label{eq:poissonized-tail-exp}
 \PP\{d(\widehat b_N,b)\ge u\mid N>0\}
 \le \min\!\left\{1,
 C_0\frac{e^{m e^{-\psi(u)}}-1}{e^m-1}
 \right\}.
\end{equation}
\end{enumerate}
\end{theorem}

\begin{proof}
Condition on \(N=n\) and use the independence of \(N\) and the mark sequence.  Equations \eqref{eq:loss-poissonization}--\eqref{eq:tail-poissonization} follow from the zero-truncated Poisson probabilities.  For \eqref{eq:poissonized-risk-bound}, apply \eqref{eq:H-expectation}.  Finally,
\[
 \mathscr P_m[(e^{-n\psi})]
 =\frac{1}{e^m-1}\sum_{n\ge1}\frac{(me^{-\psi})^n}{n!}
 =\frac{e^{m e^{-\psi}}-1}{e^m-1},
\]
which proves \eqref{eq:poissonized-tail-exp}.
\end{proof}

The theorem is deliberately modular: any valid fixed-\(n\) result can be transferred without pretending that Campbell's formula controls a random minimizer.  The next statement records the analogous asymptotic transfer.

\begin{theorem}[Random-index consistency and intrinsic CLT]
\label{thm:random-index-clt}
Let \(N_m\sim\operatorname{Poisson}(m)\) be independent of the observations.
\begin{enumerate}[label=(\roman*)]
\item If \(\widehat b_n\to b\) in probability as \(n\to\infty\), then
\[
 \widehat b_{N_m}\to b
 \quad\text{in probability conditional on }N_m>0,
 \qquad m\to\infty.
\]
\item Suppose additionally that \(\M\) is a finite-dimensional Riemannian Hadamard manifold and that, in the tangent space \(T_b\M\),
\begin{equation}
\label{eq:fixed-n-clt}
 \sqrt n\,\Log_b(\widehat b_n)\Rightarrow Z
\end{equation}
for some random vector \(Z\).  Then
\begin{equation}
\label{eq:poissonized-clt}
 \sqrt m\,\Log_b(\widehat b_{N_m})\Rightarrow Z
 \quad\text{conditional on }N_m>0.
\end{equation}
\end{enumerate}
\end{theorem}

\begin{proof}
For consistency, fix \(\varepsilon>0\) and split the conditional sum in \eqref{eq:tail-poissonization} at a deterministic \(n_0\).  The fixed-sample tail is uniformly small for \(n\ge n_0\), while \(\PP(N_m<n_0\mid N_m>0)\to0\).

For the second assertion, \(N_m/m\to1\) in probability and \(N_m\to\infty\) in probability.  A random-index version of \eqref{eq:fixed-n-clt} gives
\(\sqrt{N_m}\Log_b(\widehat b_{N_m})\Rightarrow Z\); multiplying by \(\sqrt{m/N_m}\to1\) and applying Slutsky's theorem yields \eqref{eq:poissonized-clt}.  This is the classical Anscombe principle \cite{anscombe1952} applied in the tangent space.
\end{proof}

\section{Exact laws for geodesic-supported marks}
\label{sec:geodesic}

The general empirical barycenter is nonlinear.  Exact finite-sample formulas become available when the mark law is supported on one complete geodesic, a condition that still includes nontrivial covariance variation on Gaussian manifolds.

Let \(I\subseteq\R\) be a nonempty closed interval, possibly unbounded, and let \(\gamma:I\to\M\) be an isometric geodesic parameterization, so
\begin{equation}
\label{eq:unit-geodesic}
 d(\gamma(u),\gamma(v))=|u-v|,
 \qquad u,v\in I.
\end{equation}
A complete geodesic line corresponds to \(I=\R\), while a finite geodesic segment is also allowed.

\begin{lemma}[Barycenter on a geodesic]
\label{lem:geodesic-barycenter}
Let \(Y\) be an \(I\)-valued random variable with \(\EE Y^2<\infty\), and set \(P=\gamma(Y)\).  Then
\begin{equation}
\label{eq:geodesic-pop-bary}
 b(\mathcal L(P))=\gamma(\theta),
 \qquad \theta:=\EE Y\in I.
\end{equation}
For observations \(P_i=\gamma(Y_i)\),
\begin{equation}
\label{eq:geodesic-emp-bary}
 \widehat b_n=\gamma(\overline Y_n),
 \qquad \overline Y_n:=\frac1n\sum_{i=1}^nY_i\in I.
\end{equation}
\end{lemma}

\begin{proof}
The image \(\gamma(I)\) is a closed convex subset of a Hadamard space.  Metric projection onto a closed convex subset cannot increase distance, so the minimizer of the Fr\'echet functional lies on this geodesic set.  By \eqref{eq:unit-geodesic}, the restricted functional is \(u\mapsto\EE(Y-u)^2\) on the closed convex interval \(I\).  Since both \(\theta=\EE Y\) and every empirical average belong to \(I\), the unique population and empirical minimizers are \(\theta\) and \(\overline Y_n\), respectively.
\end{proof}

\begin{theorem}[Independent geodesic marks]
\label{thm:iid-geodesic}
Let \(Y_1,Y_2,\ldots\) be independent and identically distributed with mean \(\theta\), variance \(v<\infty\), and let \(P_i=\gamma(Y_i)\).  If \(N\sim\operatorname{Poisson}(m)\) is independent, then
\begin{equation}
\label{eq:iid-geodesic-risk}
 \EE\!\left[d^2\!\left(\widehat b_N,\gamma(\theta)\right)\,\middle|\,N>0\right]
 =vH(m).
\end{equation}
If \(Y-\theta\) is \(\tau^2\)-sub-Gaussian, i.e.
\[
 \EE e^{t(Y-\theta)}\le e^{\tau^2t^2/2}\quad(t\in\R),
\]
then, for every \(u>0\),
\begin{equation}
\label{eq:iid-geodesic-tail}
 \PP\!\left\{d\!\left(\widehat b_N,\gamma(\theta)\right)\ge u\,\middle|\,N>0\right\}
 \le\min\!\left\{1,
 2\frac{e^{m\exp[-u^2/(2\tau^2)]}-1}{e^m-1}
 \right\}.
\end{equation}
\end{theorem}

\begin{proof}
By Lemma~\ref{lem:geodesic-barycenter}, conditional on \(N=n\), the squared distance is \((\overline Y_n-\theta)^2\), whose expectation is \(v/n\).  Equation \eqref{eq:iid-geodesic-risk} follows from \eqref{eq:H-expectation}.  The standard sub-Gaussian sample-mean bound is
\(\PP(|\overline Y_n-\theta|\ge u)\le2e^{-nu^2/(2\tau^2)}\).  Apply \eqref{eq:poissonized-tail-exp}.
\end{proof}

\begin{corollary}[Two arbitrary states]
\label{cor:two-state}
Let \(p_0,p_1\in\M\), \(D=d(p_0,p_1)\), and let a mark equal \(p_1\) with probability \(q\in[0,1]\) and \(p_0\) otherwise.  Then
\begin{equation}
\label{eq:two-state-bary}
 b=p_0\#_q p_1:=\gamma_{p_0,p_1}(q),
\end{equation}
and, for a Poisson sample of mean \(m\),
\begin{equation}
\label{eq:two-state-risk}
 \EE[d^2(\widehat b_N,b)\mid N>0]
 =D^2q(1-q)H(m).
\end{equation}
\end{corollary}

\begin{proof}
Use the unit-speed coordinate \(Y=DZ\), where \(Z\sim\operatorname{Bernoulli}(q)\), and apply Theorem~\ref{thm:iid-geodesic}.
\end{proof}

Corollary~\ref{cor:two-state} is not a common-covariance calculation.  In \cref{sec:gaussian}, \(p_0\) and \(p_1\) will be Gaussian distributions with distinct, possibly noncommuting, covariance matrices.

\section{Spatially correlated geodesic fields}
\label{sec:correlated}

Independent marking reduces spatial randomness to a random count.  A stronger interaction arises when the marks are samples of a common random field.  Let \(Y=\{Y(x):x\in B\}\) be an \(I\)-valued second-order random field independent of the ground Poisson process.  Define
\begin{equation}
\label{eq:field-mean-cov}
 \mu(x):=\EE Y(x),
 \qquad
 C(x,y):=\operatorname{Cov}(Y(x),Y(y)),
\end{equation}
and attach the manifold-valued mark
\begin{equation}
\label{eq:geodesic-field-mark}
 P_x:=\gamma(Y(x))
\end{equation}
to every Poisson point.  Conditional on the point locations, these marks are generally dependent.

Let \(w_B\) be the normalized intensity law in \eqref{eq:w-and-QB} and define
\begin{align}
\label{eq:theta-field}
 \theta_B&:=\int_B\mu(x)\,w_B(\mathrm dx),\\
\label{eq:VB}
 V_B&:=\int_B\left[C(x,x)+(\mu(x)-\theta_B)^2\right]w_B(\mathrm dx),\\
\label{eq:CB}
 C_B&:=\int_B\int_B C(x,y)\,w_B(\mathrm dx)w_B(\mathrm dy).
\end{align}
Assume these quantities are finite.  The natural population target is \(b_B=\gamma(\theta_B)\).

\begin{theorem}[Exact correlation decomposition]
\label{thm:correlated-field}
Let \(N_B\sim\operatorname{Poisson}(m_B)\) and let \(\widehat b_B\) be the empirical barycenter of the field marks in \(B\), conditional on \(N_B>0\).  Then
\begin{equation}
\label{eq:correlated-field-risk}
 \EE\!\left[d^2(\widehat b_B,b_B)\,\middle|\,N_B>0\right]
 =C_B+(V_B-C_B)H(m_B).
\end{equation}
More precisely, conditional on \(N_B=n\ge1\),
\begin{equation}
\label{eq:correlated-fixed-n}
 \EE[d^2(\widehat b_B,b_B)\mid N_B=n]
 =C_B+\frac{V_B-C_B}{n}.
\end{equation}
\end{theorem}

\begin{proof}
Conditional on \(N_B=n\), the locations \(X_1,\ldots,X_n\) are independent with law \(w_B\), while the marks are \(\gamma(Y(X_i))\).  Lemma~\ref{lem:geodesic-barycenter} gives
\[
 d^2(\widehat b_B,b_B)
 =\left(\frac1n\sum_{i=1}^nY(X_i)-\theta_B\right)^2.
\]
A diagonal term has expectation \(V_B\).  For \(i\ne j\), independence of \(X_i,X_j\) and the definition of \(\theta_B\) give
\begin{align*}
 &\EE[(Y(X_i)-\theta_B)(Y(X_j)-\theta_B)]\\
 &\quad=\int_B\int_B\left[C(x,y)+(\mu(x)-\theta_B)(\mu(y)-\theta_B)\right]
 w_B(\mathrm dx)w_B(\mathrm dy)=C_B.
\end{align*}
There are \(n\) diagonal and \(n(n-1)\) off-diagonal terms, yielding \eqref{eq:correlated-fixed-n}.  Averaging \(1/n\) over the zero-truncated Poisson law gives \eqref{eq:correlated-field-risk}.
\end{proof}

\subsection{Stationary fields: densification and window growth are different}

Suppose now that \(\Phi\) is homogeneous with intensity \(\lambda>0\), so \(m_B=\lambda|B|\), and that \(Y\) is second-order stationary with constant mean \(\theta\) and covariance \(C(x,y)=C_0(x-y)\).  Then
\begin{equation}
\label{eq:stationary-CB}
 V_B=C_0(0),
 \qquad
 C_B=\frac1{|B|^2}\int_B\int_B C_0(x-y)\,\mathrm dx\,\mathrm dy.
\end{equation}

\begin{corollary}[Correlation floor under densification]
\label{cor:densification-floor}
For fixed \(B\),
\begin{equation}
\label{eq:density-limit}
 \lim_{\lambda\to\infty}
 \EE[d^2(\widehat b_B,\gamma(\theta))\mid N_B>0]
 =C_B.
\end{equation}
Thus increasing the point density cannot eliminate the randomness of the spatial average of the underlying field.
\end{corollary}

\begin{proof}
Use \eqref{eq:correlated-field-risk} and \(H(\lambda|B|)\to0\).
\end{proof}

For growing windows, let \((B_R)\) be a Van Hove sequence: \(|B_R|\to\infty\) and boundary effects vanish under every fixed translation.  If \(C_0\in L^1(\R^s)\), then
\begin{equation}
\label{eq:CB-vanhove}
 |B_R|C_{B_R}\longrightarrow \int_{\R^s}C_0(h)\,\mathrm dh.
\end{equation}
Indeed,
\[
 |B_R|C_{B_R}
 =\int_{\R^s}C_0(h)
 \frac{|B_R\cap(B_R-h)|}{|B_R|}\,\mathrm dh,
\]
and dominated convergence applies.

\begin{corollary}[Large-window mean-square constant]
\label{cor:large-window}
Under the stationary assumptions above, if \(C_0\in L^1(\R^s)\), then, for fixed \(\lambda>0\),
\begin{equation}
\label{eq:large-window-limit}
 |B_R|\,
 \EE[d^2(\widehat b_{B_R},\gamma(\theta))\mid N_{B_R}>0]
 \longrightarrow
 \int_{\R^s}C_0(h)\,\mathrm dh+\frac{C_0(0)}{\lambda}.
\end{equation}
\end{corollary}

\begin{proof}
Insert \eqref{eq:stationary-CB} into \eqref{eq:correlated-field-risk}.  Use \eqref{eq:CB-vanhove}, \(|B_R|H(\lambda|B_R|)\to1/\lambda\), and \(C_{B_R}H(\lambda|B_R|)=o(|B_R|^{-1})\).
\end{proof}

The same constant governs the limiting distribution when the field is Gaussian.  This is not a random-index version of an independent-sample central limit theorem: the off-diagonal field covariance contributes at the same order as the Poisson diagonal term.

\begin{theorem}[Spatial central limit theorem for a Gaussian geodesic field]
\label{thm:spatial-gaussian-clt}
Let \(Y\) be a jointly measurable stationary Gaussian field on \(\R^s\), independent of a homogeneous Poisson process of intensity \(\lambda>0\), with mean \(\theta\) and covariance \(C_0\).  Let \((B_R)\) be a Van Hove sequence and assume
\begin{equation}
\label{eq:clt-cov-assumption}
 C_0\in L^1(\R^s)\cap L^2(\R^s).
\end{equation}
Write
\[
 \overline Y_R:=\frac1{N_{B_R}}\sum_{x\in\Phi\cap B_R}Y(x)
\]
on \(\{N_{B_R}>0\}\).  Then, conditionally on \(N_{B_R}>0\),
\begin{equation}
\label{eq:spatial-gaussian-clt}
 \sqrt{|B_R|}\,(\overline Y_R-\theta)
 \Rightarrow
 \mathcal N(0,\sigma_\lambda^2),
 \qquad
 \sigma_\lambda^2:=\int_{\R^s}C_0(h)\,\mathrm dh+\frac{C_0(0)}{\lambda}.
\end{equation}
Consequently, for the manifold-valued marks \(P_x=\gamma(Y(x))\), the signed arclength coordinate of
\(\sqrt{|B_R|}\,\Log_{\gamma(\theta)}(\widehat b_{B_R})\) has the limit in \eqref{eq:spatial-gaussian-clt} whenever \(\M\) is Riemannian.
\end{theorem}

\begin{proof}
Conditionally on the Poisson configuration, \(\overline Y_R-\theta\) is Gaussian with variance
\[
 Q_R:=\frac{\Xi_R}{N_{B_R}^2},
 \qquad
 \Xi_R:=\sum_{x,y\in\Phi\cap B_R}C_0(x-y).
\]
We first show
\begin{equation}
\label{eq:Xi-lln}
 \frac{\Xi_R}{|B_R|}
 \longrightarrow
 \lambda C_0(0)+\lambda^2\int_{\R^s}C_0(h)\,\mathrm dh
 \quad\text{in }L^2.
\end{equation}
The expectation converges to the right-hand side by the first- and second-order Campbell formulas and the Van Hove identity in \eqref{eq:CB-vanhove}.  For the variance, put
\[
 f_R(x,y):=\mathbf1_{B_R}(x)\mathbf1_{B_R}(y)C_0(x-y),
 \qquad
 U_R:=\sum_{(x,y)\in(\Phi\cap B_R)^2_{\ne}}C_0(x-y),
\]
so \(\Xi_R=C_0(0)N_{B_R}+U_R\).  The variance formula for an order-two Poisson U-statistic gives \cite{lastpenrose2017}
\begin{align*}
 \Var(U_R)
 &=2\lambda^2\int f_R(x,y)^2\,\mathrm dx\,\mathrm dy\\
 &\quad+4\lambda^3\int_{B_R}\left(\int_{B_R}C_0(x-y)\,\mathrm dy\right)^2\mathrm dx,
\end{align*}
and
\[
 \Cov(N_{B_R},U_R)=2\lambda^2\int f_R(x,y)\,\mathrm dx\,\mathrm dy.
\]
By \eqref{eq:clt-cov-assumption}, the absolute values of these three integrals are bounded respectively by
\(|B_R|\|C_0\|_2^2\), \(|B_R|\|C_0\|_1^2\), and \(|B_R|\|C_0\|_1\).  Hence \(\Var(\Xi_R)=O(|B_R|)\), proving \eqref{eq:Xi-lln}.

Also \(N_{B_R}/|B_R|\to\lambda\) in probability.  Therefore
\[
 |B_R|Q_R
 =\frac{\Xi_R/|B_R|}{(N_{B_R}/|B_R|)^2}
 \longrightarrow \sigma_\lambda^2
 \quad\text{in probability}.
\]
For every \(t\in\R\), the conditional characteristic function is
\[
 \EE\!\left[e^{it\sqrt{|B_R|}(\overline Y_R-\theta)}\,\middle|\,\Phi\right]
 =\exp\!\left(-\frac{t^2}{2}|B_R|Q_R\right).
\]
The right-hand side is bounded by one and converges in probability to
\(e^{-t^2\sigma_\lambda^2/2}\); taking expectations proves \eqref{eq:spatial-gaussian-clt}.  Finally, \(\PP(N_{B_R}=0)=e^{-\lambda|B_R|}\to0\), so zero truncation does not change the limit.
\end{proof}

Equation \eqref{eq:large-window-limit} separates the cost of the field's long-range spatial dependence from the additional variance caused by Poisson sampling at finite intensity.  The two terms have different design implications: increasing \(\lambda\) only suppresses the second term, whereas increasing the spatial aperture suppresses both through the common \(|B_R|^{-1}\) rate.

\section{Palm sampling and Poisson thinning}
\label{sec:palm-thinning}

\subsection{Reduced versus non-reduced Palm barycenters}

Let \(\Phi\) be a homogeneous Poisson process of intensity \(\lambda\), and let \(Y\) be a stationary field independent of \(\Phi\), with mean \(\theta\), variance \(v:=C_0(0)\), and covariance \(C_0(h)\).  Let \(B\) be a bounded window containing the origin and \(m=\lambda|B|\).

Under the reduced Palm distribution at the origin, Slivnyak's theorem states that the remaining process has exactly the ordinary Poisson law \cite{lastpenrose2017,chiu2013}.  Therefore a reduced-Palm barycenter that excludes the point at the origin has the same distribution as the ordinary estimator; no size-biased summation identity is needed.

A different and nontrivial estimator is obtained under the non-reduced Palm distribution by including the typical point.  Let \(N\sim\operatorname{Poisson}(m)\) be the number of other points in \(B\), and define
\begin{equation}
\label{eq:palm-aug-estimator}
 \widetilde b_B^0
 :=\gamma\!\left(
 \frac{Y(0)+\sum_{i=1}^{N}Y(X_i)}{N+1}
 \right).
\end{equation}
Set
\begin{equation}
\label{eq:a-c-palm}
 a_B:=\frac1{|B|}\int_B C_0(x)\,\mathrm dx,
 \qquad
 c_B:=\frac1{|B|^2}\int_B\int_B C_0(x-y)\,\mathrm dx\,\mathrm dy.
\end{equation}

\begin{theorem}[Exact non-reduced Palm risk]
\label{thm:palm-risk}
For the estimator in \eqref{eq:palm-aug-estimator},
\begin{align}
\label{eq:palm-risk}
 &\EE^0\!\left[d^2\!\left(\widetilde b_B^0,\gamma(\theta)\right)\right]\\
 &\quad=vG_1(m)
 +2a_B\bigl(G_1(m)-G_2(m)\bigr)
 +c_B\bigl(1-3G_1(m)+2G_2(m)\bigr),\nonumber
\end{align}
where \(G_1,G_2\) are given in \eqref{eq:G1-G2}.  If the typical mark and all other marks are independent with common coordinate variance \(v\), then
\begin{equation}
\label{eq:palm-iid}
 \EE^0[d^2(\widetilde b_B^0,\gamma(\theta))]=vG_1(m).
\end{equation}
\end{theorem}

\begin{proof}
Slivnyak's theorem gives an independent ordinary Poisson process of other points, while the independent stationary field retains its original law.  Conditional on \(N=n\), the coordinate average in \eqref{eq:palm-aug-estimator} has variance
\begin{equation}
\label{eq:palm-fixed-n}
 \frac{(n+1)v+2na_B+n(n-1)c_B}{(n+1)^2}.
\end{equation}
The three Poisson expectations are
\begin{align*}
 \EE\frac1{N+1}&=G_1,\\
 \EE\frac{N}{(N+1)^2}&=G_1-G_2,\\
 \EE\frac{N(N-1)}{(N+1)^2}&=1-3G_1+2G_2.
\end{align*}
Averaging \eqref{eq:palm-fixed-n} proves \eqref{eq:palm-risk}.  Independent marks have zero cross-covariances, yielding \eqref{eq:palm-iid}.
\end{proof}

The distinction between \(H(m)\) and \(G_1(m)\) is structural.  The ordinary estimator is conditioned on a positive random count and has effective factor \(H(m)\); the non-reduced Palm estimator always contains the typical point and has total sample size \(N+1\).

\subsection{Exact distortion under independent thinning}

Return to the geodesic random-field model of \cref{sec:correlated}, with finite \(V_B\) and \(C_B\).  Let the full Poisson sample in \(B\) have mean count \(m\), and retain each point independently with probability \(p\in(0,1]\).  By Poisson splitting, the retained and discarded point processes are independent conditional on the common field, and their counts are independent,
\begin{equation}
\label{eq:split-counts}
 K\sim\operatorname{Poisson}(pm),
 \qquad
 L\sim\operatorname{Poisson}((1-p)m).
\end{equation}
Let \(\widehat b_{\mathrm{all}}\) be the barycenter of all \(K+L\) marks and \(\widehat b_{\mathrm{ret}}\) the barycenter of the retained \(K\) marks.  We condition on \(K>0\), which also guarantees \(K+L>0\).

Define
\begin{equation}
\label{eq:J-def}
 J(m,p):=
 \frac{e^{-m}\bigl[A(m)-A((1-p)m)\bigr]}{1-e^{-pm}},
\end{equation}
where \(A\) is given in \eqref{eq:A-H-def}.  Equivalently,
\begin{equation}
\label{eq:J-Ei}
 J(m,p)=\frac{e^{-m}}{1-e^{-pm}}
 \left[
 \operatorname{Ei}(m)-\operatorname{Ei}((1-p)m)
 -\log\frac1{1-p}
 \right]
\end{equation}
for \(p\in(0,1)\), with the continuous limit used at \(p=1\).

\begin{theorem}[Poisson thinning distortion for a geodesic field]
\label{thm:thinning}
The expected squared displacement between the retained and full barycenters is
\begin{equation}
\label{eq:thinning-risk}
 \EE\!\left[d^2(\widehat b_{\mathrm{ret}},\widehat b_{\mathrm{all}})\,\middle|\,K>0\right]
 =(V_B-C_B)\bigl[H(pm)-J(m,p)\bigr].
\end{equation}
As \(m\to\infty\) with fixed \(p\in(0,1]\),
\begin{equation}
\label{eq:thinning-asymptotic}
 \EE[d^2(\widehat b_{\mathrm{ret}},\widehat b_{\mathrm{all}})\mid K>0]
 =\frac{(V_B-C_B)(1-p)}{pm}+O(m^{-2}).
\end{equation}
In particular, the correlation floor \(C_B\) cancels because the retained and full barycenters are computed from the same field realization.
\end{theorem}

\begin{proof}
Conditional on \((K,L)=(k,\ell)\) with \(k\ge1\), write the retained and discarded coordinate means as \(\overline Y_K\) and \(\overline Y_L\).  If \(\ell\ge1\), the calculation in Theorem~\ref{thm:correlated-field} gives
\[
 \Var(\overline Y_K)=C_B+\frac{V_B-C_B}{k},\qquad
 \Var(\overline Y_L)=C_B+\frac{V_B-C_B}{\ell},\qquad
 \Cov(\overline Y_K,\overline Y_L)=C_B.
\]
Since
\(
 \overline Y_{K+L}=(k\overline Y_K+\ell\overline Y_L)/(k+\ell)
\), we obtain, with the value zero when \(\ell=0\),
\begin{equation}
\label{eq:conditional-thin}
 \EE\!\left[(\overline Y_K-\overline Y_{K+L})^2\,\middle|\,K=k,L=\ell\right]
 =(V_B-C_B)\frac{\ell}{k(k+\ell)}
 =(V_B-C_B)\left(\frac1k-\frac1{k+\ell}\right).
\end{equation}
The first expectation in \eqref{eq:conditional-thin} averages to \(H(pm)\).  For the second,
\begin{align*}
 &\EE\!\left[\frac{\mathbf1_{\{K>0\}}}{K+L}\right]\\
 &\quad=e^{-m}A(m)-e^{-m}A((1-p)m),
\end{align*}
where the second term subtracts the configurations with \(K=0,L>0\).  Division by \(\PP(K>0)=1-e^{-pm}\) yields \(J(m,p)\), proving \eqref{eq:thinning-risk}.  Expansion of \(H(pm)\) and \(J(m,p)\) gives \eqref{eq:thinning-asymptotic}.
\end{proof}

For comparison, two independent point-process resamples need not produce independent barycenters when they interrogate the same random field.

\begin{corollary}[Independent Poisson resamples]
\label{cor:independent-resamples}
Let two independent Poisson processes of mean counts \(m_1,m_2\) sample the same geodesic field in \(B\), and condition on both counts being positive.  Then
\begin{equation}
\label{eq:same-field-resample-risk}
 \EE[d^2(\widehat b^{(1)},\widehat b^{(2)})\mid N_1>0,N_2>0]
 =(V_B-C_B)\bigl[H(m_1)+H(m_2)\bigr].
\end{equation}
If the two processes instead sample independent copies of the field, then
\begin{equation}
\label{eq:independent-field-resample-risk}
 \EE[d^2(\widehat b^{(1)},\widehat b^{(2)})\mid N_1>0,N_2>0]
 =2C_B+(V_B-C_B)\bigl[H(m_1)+H(m_2)\bigr].
\end{equation}
For independent identically distributed marks, \(C_B=0\) and \(V_B=v\), so both cases reduce to
\begin{equation}
\label{eq:independent-resample-risk}
 v\bigl[H(m_1)+H(m_2)\bigr].
\end{equation}
\end{corollary}

\begin{proof}
Each sample mean has variance \(C_B+(V_B-C_B)H(m_i)\).  For two location resamples of the same field, their covariance is \(C_B\); for independent field copies it is zero.  Subtracting twice the relevant covariance gives \eqref{eq:same-field-resample-risk} and \eqref{eq:independent-field-resample-risk}.
\end{proof}

\section{Gaussian information-geometric specializations}
\label{sec:gaussian}

The abstract results become information-geometric once \(\M\) is chosen as a Gaussian statistical manifold.  We emphasize models that retain covariance variation and possess a globally nonpositive geometry.

\subsection{The full univariate Fisher--Rao Gaussian manifold}
\label{subsec:univariate-fr}

Parameterize a univariate normal distribution by \(p=(\mu,\sigma)\equiv\mathcal N(\mu,\sigma^2)\), with \(\sigma>0\).  Its Fisher--Rao line element is \cite{amari2000,costa2015,miyamoto2024}
\begin{equation}
\label{eq:univ-fr-metric}
 \mathrm ds_{\mathrm{FR}}^2
 =\frac{\mathrm d\mu^2+2\,\mathrm d\sigma^2}{\sigma^2}.
\end{equation}
With \(x=\mu/\sqrt2\) and \(y=\sigma\), \eqref{eq:univ-fr-metric} is twice the Poincar\'e half-plane metric.  Hence the manifold is Hadamard with constant sectional curvature \(-1/2\), and
\begin{equation}
\label{eq:univ-fr-distance}
 d_{\mathrm{FR}}(p_1,p_2)
 =\sqrt2\,\operatorname{arcosh}\!\left(
 1+\frac{(\mu_1-\mu_2)^2+2(\sigma_1-\sigma_2)^2}
 {4\sigma_1\sigma_2}
 \right).
\end{equation}

Equal endpoint variances do not imply a constant-variance geodesic.  The next proposition records the simplest counterexample.

\begin{proposition}[Midpoint of a symmetric equal-variance pair]
\label{prop:fr-midpoint}
Let
\[
 p_-:=\mathcal N(-a,\sigma^2),
 \qquad
 p_+:=\mathcal N(a,\sigma^2),
 \qquad a\ne0.
\]
Their unique Fisher--Rao geodesic midpoint in the full manifold is
\begin{equation}
\label{eq:fr-midpoint}
 p_{1/2}=\mathcal N\!\left(0,\sigma^2+\frac{a^2}{2}\right).
\end{equation}
In particular, \(\mathcal N(0,\sigma^2)\) is only the midpoint of the artificially restricted fixed-variance submanifold.
\end{proposition}

\begin{proof}
In Poincar\'e coordinates, the endpoints are \((\pm a/\sqrt2,\sigma)\).  Their geodesic is the semicircle orthogonal to the boundary and centered at the origin, with Euclidean radius \(r=\sqrt{\sigma^2+a^2/2}\).  Symmetry places the hyperbolic midpoint at the top of the semicircle, \((0,r)\), proving \eqref{eq:fr-midpoint}.
\end{proof}

\begin{figure}[t]
\centering
\includegraphics[width=0.96\linewidth]{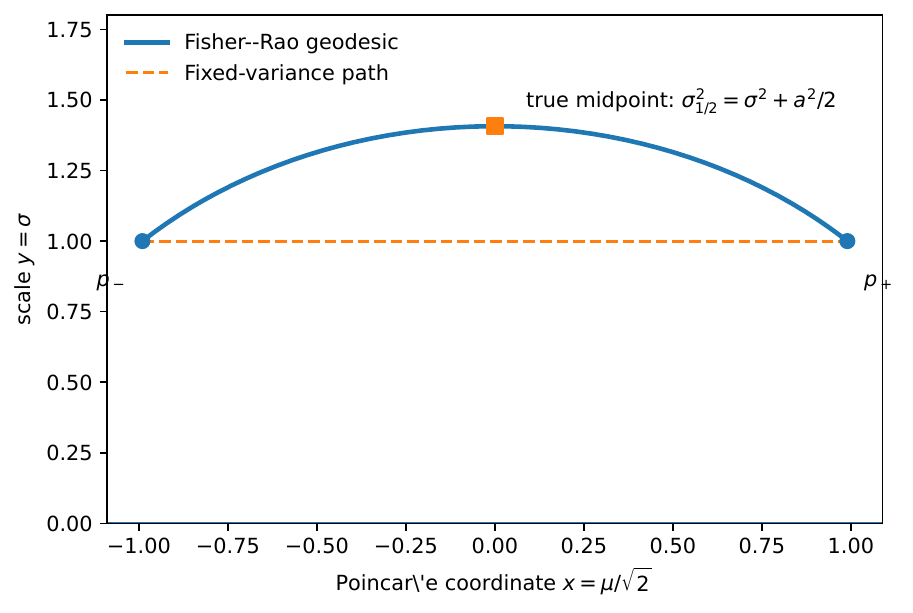}
\caption{Two equal-variance univariate Gaussians and their full Fisher--Rao geodesic.  The true midpoint has variance \(\sigma^2+a^2/2\); the horizontal fixed-variance segment is not a geodesic of the full manifold.}
\label{fig:fr-midpoint}
\end{figure}

\Cref{fig:fr-midpoint} illustrates the distinction.  The full-manifold distance between equal-variance Gaussians is obtained from \eqref{eq:univ-fr-distance}, not from the Mahalanobis length of the horizontal constant-variance path.

A nontrivial but exactly tractable submodel fixes the mean and allows the scale to vary.  Along the vertical geodesic \(\mu=\mu_0\),
\begin{equation}
\label{eq:vertical-fr-distance}
 d_{\mathrm{FR}}\!\left(\mathcal N(\mu_0,\sigma_1^2),
 \mathcal N(\mu_0,\sigma_2^2)\right)
 =\sqrt2\,|\log(\sigma_2/\sigma_1)|.
\end{equation}
Thus
\begin{equation}
\label{eq:logscale-coordinate}
 Y=\sqrt2\log(\sigma/\sigma_0)
\end{equation}
is a unit-speed geodesic coordinate.

\begin{corollary}[Poisson-sampled log-scale Gaussian field]
\label{cor:logscale-field}
Let \(Y(x)\) be a second-order random field and attach
\begin{equation}
\label{eq:logscale-mark}
 P_x=\mathcal N\!\left(\mu_0,
 \sigma_0^2e^{\sqrt2Y(x)}\right).
\end{equation}
Then the window population barycenter is
\begin{equation}
\label{eq:logscale-pop}
 b_B=\mathcal N\!\left(\mu_0,
 \sigma_0^2e^{\sqrt2\theta_B}\right),
\end{equation}
where \(\theta_B\) is defined in \eqref{eq:theta-field}, and the exact Fisher--Rao mean-square error is \eqref{eq:correlated-field-risk}.  In the independent-mark case with \(\operatorname{Var}Y=v\), it reduces to \(vH(m_B)\).
\end{corollary}

\begin{proof}
Equations \eqref{eq:vertical-fr-distance}--\eqref{eq:logscale-coordinate} identify \(Y\) as the arclength coordinate, so Lemma~\ref{lem:geodesic-barycenter} and Theorem~\ref{thm:correlated-field} apply directly.
\end{proof}

More generally, any two univariate Gaussians define a unique full-manifold geodesic.  Therefore Corollary~\ref{cor:two-state} gives an exact random-count risk for a two-regime field with simultaneous mean and variance changes, using \(D\) from \eqref{eq:univ-fr-distance}.

\subsection{Multivariate covariance geometry with fixed mean}
\label{subsec:spd-fr}

Let \(\Spp{d}\) denote the cone of \(d\times d\) symmetric positive-definite matrices, and fix \(\mu_0\in\R^d\).  The Gaussian submanifold
\begin{equation}
\label{eq:fixed-mean-manifold}
 \mathcal G_{\mu_0}:=
 \{\mathcal N(\mu_0,\Sigma):\Sigma\in\Spp{d}\}
\end{equation}
has Fisher metric
\begin{equation}
\label{eq:spd-fr-metric}
 g_\Sigma(U,V)=\frac12\tr(\Sigma^{-1}U\Sigma^{-1}V).
\end{equation}
It is a Riemannian Hadamard manifold.  Its distance and geodesic are \cite{bhatia2007,moakher2005,pinele2020}
\begin{align}
\label{eq:spd-fr-distance}
 d_{\mathrm{FR}}(\Sigma_0,\Sigma_1)
 &=\frac1{\sqrt2}
 \left\|\log\!\left(\Sigma_0^{-1/2}\Sigma_1\Sigma_0^{-1/2}\right)\right\|_{\mathrm F},\\
\label{eq:spd-geodesic}
 \Sigma_0\#_t\Sigma_1
 &:=\Sigma_0^{1/2}
 \left(\Sigma_0^{-1/2}\Sigma_1\Sigma_0^{-1/2}\right)^t
 \Sigma_0^{1/2},
 \qquad t\in\R.
\end{align}
No commutativity assumption is required.

Let \(Q\) be a probability law on \(\Spp{d}\) with finite second Fisher moment, let \(\overline\Sigma=b(Q)\), and let \(\widehat\Sigma_n\) be the affine-invariant empirical barycenter of independent draws \(S_1,\ldots,S_n\sim Q\).  The Karcher equation is
\begin{equation}
\label{eq:karcher-equation}
 \int_{\Spp{d}}\Log_{\overline\Sigma}(S)\,Q(\mathrm dS)=0,
\end{equation}
where
\[
 \Log_\Sigma(S)
 =\Sigma^{1/2}\log\!\left(\Sigma^{-1/2}S\Sigma^{-1/2}\right)\Sigma^{1/2}.
\]
Unlike an arithmetic or precision average, \eqref{eq:karcher-equation} remains nonlinear for general noncommuting matrices.

\begin{proposition}[Marked-Poisson covariance barycenter and intrinsic limit]
\label{prop:spd-clt}
Let \(B\subset\R^s\) be a bounded window with normalized location law \(w_B\).  At location \(x\), let the covariance mark have a measurable law \(Q_x\) on \(\Spp{d}\), and assume
\[
 \int_B\int_{\Spp{d}}
 d_{\mathrm{FR}}^2(S,\Sigma_\circ)\,Q_x(\mathrm dS)w_B(\mathrm dx)<\infty
\]
for one, and hence every, \(\Sigma_\circ\in\Spp{d}\).  Define the mixture law \(Q_B=\int_BQ_xw_B(\mathrm dx)\).  Then the population covariance barycenter \(\overline\Sigma_B=b(Q_B)\) exists uniquely and is characterized by the noncommutative Karcher equation
\begin{equation}
\label{eq:spatial-karcher}
 \int_B\int_{\Spp{d}}
 \Log_{\overline\Sigma_B}(S)\,Q_x(\mathrm dS)w_B(\mathrm dx)=0.
\end{equation}
Equivalently,
\[
 \int_B\int_{\Spp{d}}
 \log\!\left(\overline\Sigma_B^{-1/2}S\overline\Sigma_B^{-1/2}\right)
 Q_x(\mathrm dS)w_B(\mathrm dx)=0.
\]
No commutativity or simultaneous-diagonalization assumption is required.

Let the ground process have intensity measure \(t\Lambda\), and let \(\widehat\Sigma_{B,t}\) be the empirical affine-invariant barycenter, conditional on a nonempty window.  Then
\[
 \widehat\Sigma_{B,t}\longrightarrow\overline\Sigma_B
 \qquad\text{in probability as }t\to\infty.
\]
Assume additionally the standard intrinsic-CLT regularity conditions at \(\overline\Sigma_B\).  Let
\[
 \mathcal F_B(\Sigma):=\frac12\int_{\Spp{d}}d_{\mathrm{FR}}^2(\Sigma,S)Q_B(\mathrm dS),
\]
let \(\mathcal A_B\) be its nonsingular self-adjoint Hessian at \(\overline\Sigma_B\), and let \(\mathcal V_B\) be the covariance operator of \(\Log_{\overline\Sigma_B}(S)\), \(S\sim Q_B\), in the Fisher tangent inner product.  Since the mean count is \(tm_B\),
\begin{equation}
\label{eq:spd-poisson-clt}
 \sqrt{tm_B}\,\Log_{\overline\Sigma_B}(\widehat\Sigma_{B,t})
 \Rightarrow
 \mathcal N\!\left(0,\mathcal A_B^{-1}\mathcal V_B\mathcal A_B^{-1}\right).
\end{equation}
\end{proposition}

\begin{proof}
The affine-invariant covariance manifold is finite-dimensional and Hadamard.  The finite second-moment assumption therefore yields a unique barycenter, and differentiating its Fr\'echet functional gives \eqref{eq:spatial-karcher}.  Proposition~\ref{prop:conditional-iid} identifies the covariance observations, conditional on their Poisson count, as an independent sample from \(Q_B\).  Strong consistency follows from Corollary~\ref{cor:intensity-consistency}.  Under the stated smoothness and nonsingularity assumptions, the fixed-sample intrinsic central limit theorem holds on the tangent space \cite{bhattacharya2003,bhattacharya2005,pennec2006}; Theorem~\ref{thm:random-index-clt} then gives \eqref{eq:spd-poisson-clt}.
\end{proof}

\begin{corollary}[Two covariance regimes]
\label{cor:two-covariance}
Let a Gaussian mark be \(\mathcal N(\mu_0,\Sigma_1)\) with probability \(q\) and \(\mathcal N(\mu_0,\Sigma_0)\) otherwise.  Define
\begin{equation}
\label{eq:D-spd}
 D^2:=\frac12
 \left\|\log\!\left(\Sigma_0^{-1/2}\Sigma_1\Sigma_0^{-1/2}\right)\right\|_{\mathrm F}^2.
\end{equation}
Then the population Fisher--Rao barycenter is
\begin{equation}
\label{eq:two-cov-pop}
 \mathcal N(\mu_0,\Sigma_0\#_q\Sigma_1),
\end{equation}
and, for a Poisson sample of mean \(m\),
\begin{equation}
\label{eq:two-cov-risk}
 \EE[d_{\mathrm{FR}}^2(\widehat b_N,b)\mid N>0]
 =D^2q(1-q)H(m).
\end{equation}
\end{corollary}

\begin{proof}
The curve \(t\mapsto\Sigma_0\#_t\Sigma_1\) has constant speed \(D\).  Apply Corollary~\ref{cor:two-state}.
\end{proof}

The same construction accommodates a spatially correlated covariance field.  Let \(T(x)\) be a real field and set
\begin{equation}
\label{eq:cov-geodesic-field}
 \Sigma(x)=\Sigma_0\#_{T(x)}\Sigma_1.
\end{equation}
The unit-speed coordinate is \(Y(x)=DT(x)\).  Theorem~\ref{thm:correlated-field}, its densification floor, the Palm formula, the large-window limit, and---when \(T\) is Gaussian---the spatial central limit theorem therefore hold exactly after replacing the scalar covariance of \(Y\) by \(D^2\) times the covariance of \(T\).  This remains nontrivial when \(\Sigma_0\Sigma_1\ne\Sigma_1\Sigma_0\).  For example,
\[
 \Sigma_0=\begin{pmatrix}1&0\\0&4\end{pmatrix},
 \qquad
 \Sigma_1=R_{\pi/4}\begin{pmatrix}9&0\\0&1\end{pmatrix}R_{\pi/4}^{\mathsf T}
 =\begin{pmatrix}5&4\\4&5\end{pmatrix}
\]
do not commute, yet \eqref{eq:two-cov-pop}--\eqref{eq:two-cov-risk} remain exact.

\subsection{Wasserstein geometry: a boundary case, not the main model}
\label{subsec:wasserstein}

For completeness, the squared \(2\)-Wasserstein distance between Gaussian distributions is \cite{takatsu2011,malago2018}
\begin{align}
\label{eq:gaussian-W2}
 W_2^2\!\left(\mathcal N(\mu_1,\Sigma_1),
 \mathcal N(\mu_2,\Sigma_2)\right)
 &=\|\mu_1-\mu_2\|^2\\
 &\quad+\tr\!\left(\Sigma_1+\Sigma_2
 -2(\Sigma_2^{1/2}\Sigma_1\Sigma_2^{1/2})^{1/2}\right).\nonumber
\end{align}
The Gaussian Wasserstein barycenter always has the arithmetic weighted mean of the \(\mu_i\); its covariance solves a nonlinear fixed-point equation \cite{alvarez2016}.  It is not a precision-weighted mean estimator.

In one dimension, write \(\Sigma_i=\sigma_i^2\).  Since
\[
 W_2^2\!\left(\mathcal N(\mu_1,\sigma_1^2),
 \mathcal N(\mu_2,\sigma_2^2)\right)
 =|\mu_1-\mu_2|^2+|\sigma_1-\sigma_2|^2,
\]
the barycenter standard deviation and variance are
\begin{equation}
\label{eq:w2-1d-correction}
 \overline\sigma=\sum_iw_i\sigma_i,
 \qquad
 \overline\Sigma=\overline\sigma^{\,2}
 =\left(\sum_iw_i\sigma_i\right)^2.
\end{equation}
More generally, if all covariance matrices commute and share an eigenbasis \(U\), then their vectors of eigenvalue square roots form Euclidean coordinates and the barycenter is their weighted Euclidean mean.  If those coordinates are supported on a closed line segment contained in the positive orthant, Lemma~\ref{lem:geodesic-barycenter} gives the corresponding scalar random-count formulas on that segment.  The full Bures--Wasserstein covariance geometry, however, is not globally CAT(0), so the general Hadamard statements in this paper are not asserted for it.

\section{Numerical validation}
\label{sec:numerics}

All experiments are reproducible using the accompanying script \texttt{generate\_figures.py}.  The numerical work is used only to validate exact formulas; no fitted parameters or empirical performance claims enter the theorems.

\subsection{Reciprocal-count factors}

\begin{figure}[t]
\centering
\includegraphics[width=0.94\linewidth]{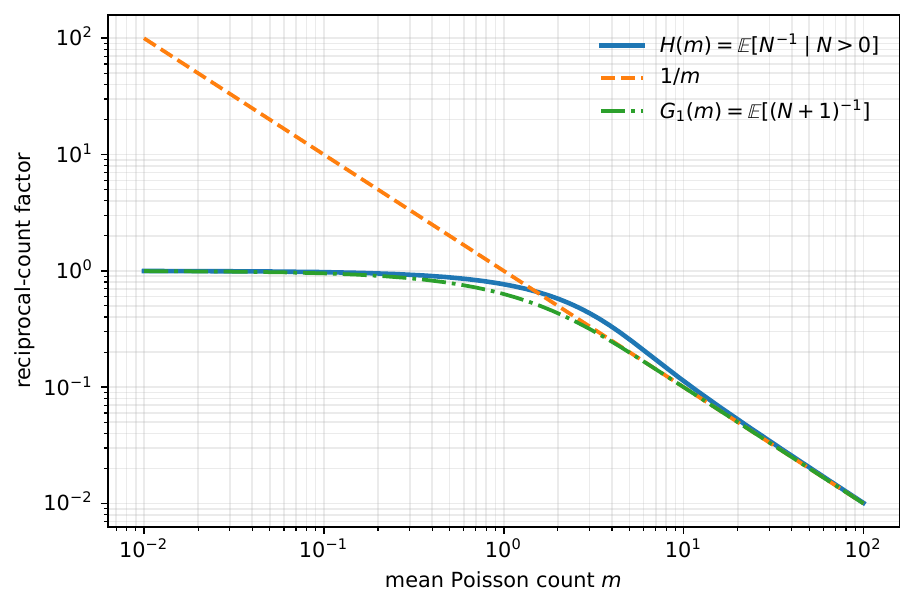}
\caption{Exact Poisson reciprocal-count factors.  The quantity relevant to an empirical barycenter conditioned on a nonempty window is \(H(m)\), not \(1/m\).  The non-reduced Palm estimator has the different factor \(G_1(m)\).}
\label{fig:count-factors}
\end{figure}

\Cref{fig:count-factors} compares \(H(m)\), the approximation \(1/m\), and the Palm factor \(G_1(m)\).  When \(m\) is small, conditioning on a nonempty window makes the count close to one, so \(H(m)\to1\), whereas \(1/m\) diverges.  At moderate and large \(m\), \(1/m\) underestimates \(H(m)\), consistently with \eqref{eq:H-asymptotic}.

\subsection{A correlated Fisher--Rao log-scale field}

Consider the univariate Gaussian field in Corollary~\ref{cor:logscale-field} on \(B=[0,L]\), with \(\mu_0=0\), \(\sigma_0=1\), and a stationary zero-mean Gaussian arclength field satisfying
\begin{equation}
\label{eq:OU-cov}
 \operatorname{Cov}(Y(x),Y(y))=v e^{-|x-y|/\ell}.
\end{equation}
The spatial-average variance in \eqref{eq:stationary-CB} is
\begin{equation}
\label{eq:OU-CB}
 C_B=\frac{2v}{L^2}\left[L\ell-\ell^2(1-e^{-L/\ell})\right].
\end{equation}
Hence the exact Fisher--Rao mean-square error is
\begin{equation}
\label{eq:OU-risk}
 C_B+(v-C_B)H(\lambda L).
\end{equation}

\begin{figure}[t]
\centering
\includegraphics[width=0.94\linewidth]{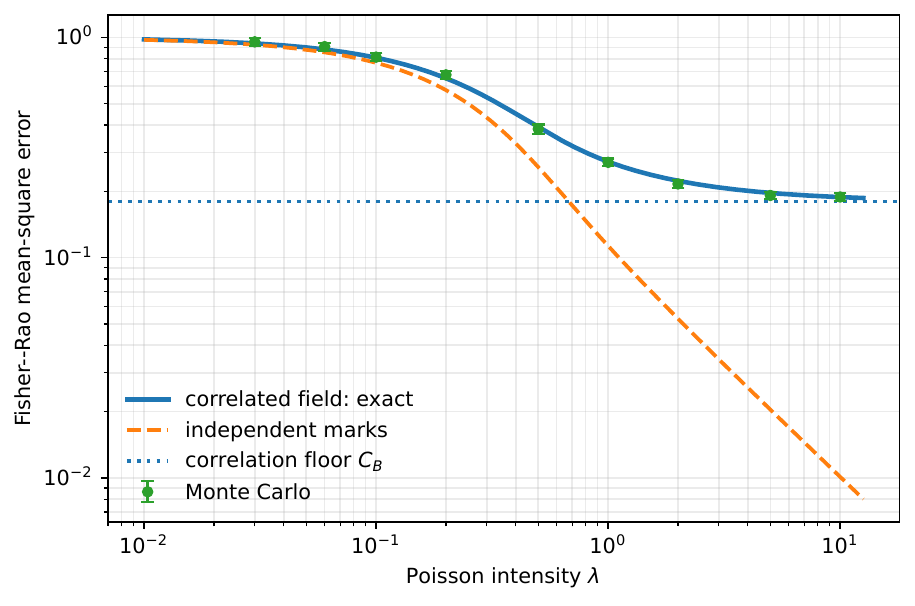}
\caption{Fisher--Rao MSE for the correlated log-scale field in \eqref{eq:OU-cov}.  Densification removes Poisson sampling noise but not the randomness of the field's spatial average.  Error bars show twice the Monte Carlo standard error.}
\label{fig:correlation-floor}
\end{figure}

\Cref{fig:correlation-floor} uses \(L=10\), \(\ell=1\), and \(v=1\).  The Monte Carlo points are based on 5000 nonempty realizations per intensity; stationary Ornstein--Uhlenbeck samples at sorted Poisson locations are generated recursively.  The simulation agrees with \eqref{eq:OU-risk}.  The independent-mark expression \(vH(\lambda L)\) decays to zero, while the correlated field converges to the nonzero floor \(C_B\).

\subsection{Thinning stability}

\begin{figure}[t]
\centering
\includegraphics[width=0.94\linewidth]{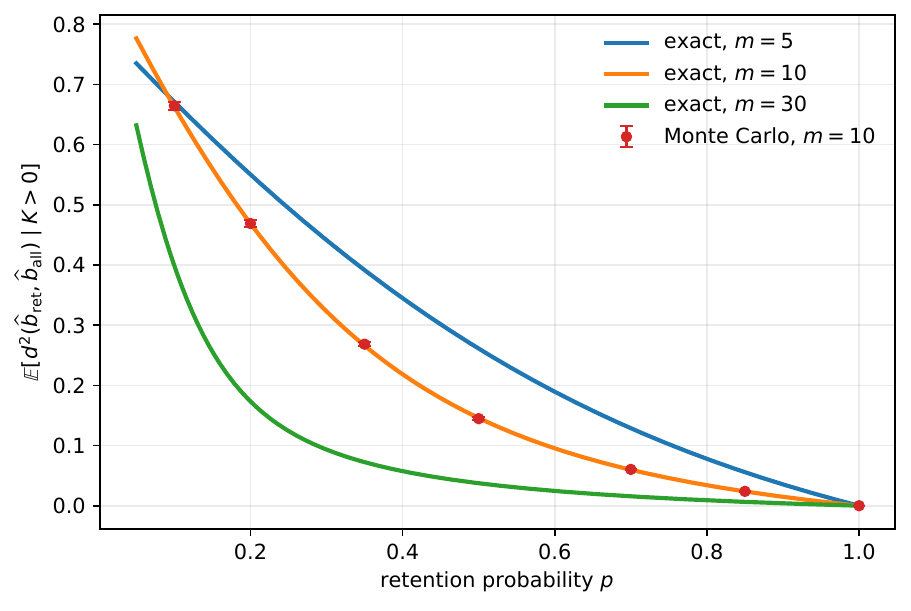}
\caption{Expected squared geodesic displacement between the retained and full barycenters under independent thinning.  Curves are the exact formula \eqref{eq:thinning-risk}; points are Monte Carlo checks for \(m=10\).}
\label{fig:thinning}
\end{figure}

\Cref{fig:thinning} evaluates \eqref{eq:thinning-risk} for unit innovation variance \(V_B-C_B=1\) and several mean counts.  Independent Monte Carlo checks for \(m=10\) use the conditional variance in \eqref{eq:conditional-thin}.  The distortion vanishes at \(p=1\), increases as more points are discarded, and approaches the large-count law \((V_B-C_B)(1-p)/(pm)\).  At finite \(m\), the exact result captures both zero truncation and the dependence between the retained and full samples.

\subsection{Noncommuting covariance regimes}

To validate the genuinely matrix-valued specialization, consider
\[
 \Sigma_0=\begin{pmatrix}1&0\\0&4\end{pmatrix},
 \qquad
 \Sigma_1=\begin{pmatrix}5&4\\4&5\end{pmatrix},
 \qquad q=0.35.
\]
The two matrices do not commute.  Their squared Fisher--Rao separation is
\[
 D^2=\frac12\left\|\log\!\left(
 \Sigma_0^{-1/2}\Sigma_1\Sigma_0^{-1/2}
 \right)\right\|_{\mathrm F}^2
 =2.02431684.
\]
For a Poisson sample, the empirical covariance barycenter lies at the random geodesic fraction \(K/N\), where \(K\mid N\sim\operatorname{Binomial}(N,q)\).  Consequently, Corollary~\ref{cor:two-covariance} gives the exact risk \(D^2q(1-q)H(m)\).  Table~\ref{tab:noncommuting} compares this expression with \(200{,}000\) zero-truncated Monte Carlo trials (seed 20260714).

\begin{table}[H]
\centering
\caption{Exact and simulated Fisher--Rao risks for two noncommuting covariance regimes.  The reported uncertainty is twice the Monte Carlo standard error.}
\label{tab:noncommuting}
\begin{tabular}{@{}rrrr@{}}
\toprule
Mean count \(m\) & Exact risk & Monte Carlo & Absolute error \\
\midrule
5  & 0.118711 & 0.118578 $\pm$ 0.000756 & 0.000133 \\
20 & 0.024315 & 0.024394 $\pm$ 0.000158 & 0.000079 \\
80 & 0.005830 & 0.005842 $\pm$ 0.000037 & 0.000011 \\
\bottomrule
\end{tabular}
\end{table}

\section{Discussion and limitations}
\label{sec:discussion}

\subsection{What the two geometries contribute}

The point-process geometry determines how many marks are observed, where they are sampled, how Palm conditioning changes the sample, and how thinning couples a reduced configuration to the original one.  The information geometry determines the intrinsic distance, the geodesics on which exact reductions are possible, and the meaning of the resulting barycenter.  Neither side can be replaced by notation alone.  In particular:

\begin{itemize}
\item With independent marks, spatial Poisson sampling enters through the exact count transform \(\mathscr P_m\), while location heterogeneity enters through the mixture law \(Q_B\).
\item With a common random field, off-diagonal covariance terms survive the location average and create the floor \(C_B\).  This is a genuinely second-order stochastic-geometric effect and cannot be recovered from Campbell's first-moment formula.
\item Under Palm conditioning, excluding the typical point gives the ordinary law by Slivnyak's theorem, whereas including it changes both the denominator and the covariance structure.
\item Under thinning, the full and retained barycenters share observations.  Their distortion is therefore not the sum of two independent estimation errors.
\end{itemize}

\subsection{Why the fixed-covariance mean-only model is excluded}

A fixed-covariance restriction can be useful when it is explicitly imposed as a statistical model.  On that restricted submanifold, Fisher length in the mean coordinate is Mahalanobis and the Fr\'echet mean is an arithmetic average.  Those facts are elementary.  They must not be interpreted as statements about the full Gaussian Fisher--Rao manifold, because the shortest path can leave the fixed-covariance set, as Proposition~\ref{prop:fr-midpoint} shows.  The covariance-varying models in \cref{subsec:univariate-fr,subsec:spd-fr} preserve the non-Euclidean structure.

\subsection{Scope of the exact correlation theory}

The exact second-order formulas require the marks to lie on one geodesic.  This condition is broader than a two-point model and includes log-scale Gaussian fields and matrix covariance fields of the form \eqref{eq:cov-geodesic-field}.  It does not cover an arbitrary correlated field wandering in several tangent directions of a curved manifold.  Extending \eqref{eq:correlated-field-risk} to such fields requires controlling curvature-dependent remainders in logarithmic coordinates or developing a genuinely manifold-valued covariance operator.  That is a natural next problem, but it is not assumed away here.

The full multivariate Gaussian Fisher--Rao manifold with simultaneous mean and covariance variation is also not treated as a Hadamard space.  Closed-form distances are unavailable in general and its curvature structure is more complicated than that of the univariate or fixed-mean covariance manifolds.  Likewise, the full Gaussian Bures--Wasserstein covariance space lies outside the CAT(0) framework.  The paper therefore states results only where the geometry needed by the proofs is valid.

\subsection{Barycenter preservation is not automatically semantic preservation}

A barycenter is a task-relevant summary only when the downstream objective depends on that summary.  Two spatial fields can have the same barycenter and different covariance functions, local extremes, or spatial arrangements.  Consequently, \eqref{eq:thinning-risk} is presented as a precise \emph{barycenter-stability} law, not as a universal semantic-fidelity guarantee.  A communication or inference application must specify which functionals of the field carry meaning and then analyze an appropriate multi-feature distortion.  Even for barycenter-only distortion, the exact random-count factor is \(H(m)\), not \(1/m\): for independent identically distributed resamples the discrepancy is \eqref{eq:independent-resample-risk}, while common-field and independent-field resampling obey the distinct laws \eqref{eq:same-field-resample-risk} and \eqref{eq:independent-field-resample-risk}.

\section{Conclusion}
\label{sec:conclusion}

We developed a rigorous finite-window framework for Fr\'echet means of distribution-valued marks sampled by a Poisson point process.  The population target is the barycenter of a normalized window mark law, and the empirical target is conditioned on a nonempty count.  This formulation leads to an exact Poissonization calculus for fixed-sample risk, tails, consistency, and intrinsic central limit theorems.

For geodesic-supported marks, we obtained exact random-count mean-square errors.  When the marks arise from a spatially correlated field, the error decomposes into a correlation floor and a Poisson term; fixed-window densification and expanding-window averaging therefore have fundamentally different limits.  For stationary Gaussian fields, the same decomposition yields a spatial central limit theorem whose variance is the sum of the integrated field covariance and the Poisson diagonal contribution.  Slivnyak's theorem yielded valid reduced- and non-reduced Palm laws, while Poisson splitting yielded an exact thinning distortion.  The specialization to the hyperbolic univariate Fisher--Rao manifold and the affine-invariant covariance manifold provides nontrivial Gaussian examples with varying covariance, including noncommuting matrix regimes.

The resulting theory is deliberately narrower than a multi-application framework, but materially stronger: every target is well-defined, zero counts are explicit, dependence is separated from independent marking, and curvature assumptions are matched to the Gaussian submanifolds on which they hold.

\appendix

\section{Count identities and asymptotic expansions}
\label{app:count-identities}

For completeness, the power-series identity
\[
 \operatorname{Ei}(m)=\gamma+\log m+\sum_{n=1}^{\infty}\frac{m^n}{n\,n!},
 \qquad m>0,
\]
gives \(A(m)\) in \eqref{eq:A-H-def}.  Therefore
\begin{align*}
 \EE[N^{-1};N>0]
 &=e^{-m}\sum_{n=1}^{\infty}\frac{m^n}{n\,n!}
 =e^{-m}A(m),\\
 \EE[N^{-1}\mid N>0]
 &=\frac{e^{-m}A(m)}{1-e^{-m}}
 =\frac{A(m)}{e^m-1}.
\end{align*}
Similarly,
\begin{align*}
 \EE\frac1{N+1}
 &=e^{-m}\sum_{n=0}^{\infty}\frac{m^n}{n!(n+1)}
 =\frac{1-e^{-m}}{m},\\
 \EE\frac1{(N+1)^2}
 &=e^{-m}\sum_{n=0}^{\infty}\frac{m^n}{n!(n+1)^2}\\
 &=\frac{e^{-m}}{m}\sum_{k=1}^{\infty}\frac{m^k}{k\,k!}
 =\frac{e^{-m}A(m)}{m}.
\end{align*}
The classical asymptotic series
\[
 \operatorname{Ei}(m)
 \sim\frac{e^m}{m}\left(1+\frac1m+\frac{2!}{m^2}+\frac{3!}{m^3}+\cdots\right)
\]
yields \eqref{eq:H-asymptotic}.

\section{A corrected Gaussian quadratic-form tail}
\label{app:quadratic-tail}

Although fixed-covariance mean-only models are not central to this paper, it is useful to record the correct tail form for the corresponding quadratic distance.  Let \(Z\sim\mathcal N(0,\Gamma)\), let \(A\succeq0\), and define
\[
 X:=Z^\top A Z,
 \qquad
 M:=\Gamma^{1/2}A\Gamma^{1/2}.
\]
Then \(\EE X=\tr M\), and the Laurent--Massart/Hsu--Kakade--Zhang inequality gives, for every \(t>0\),
\begin{equation}
\label{eq:correct-quadratic-tail}
 \PP\!\left\{
 X\ge\tr M+2\sqrt{\tr(M^2)t}+2\|M\|_{\mathrm op}t
 \right\}\le e^{-t}.
\end{equation}
See \cite{laurent2000,hsu2012}.  The operator-norm term in \eqref{eq:correct-quadratic-tail} produces the necessary sub-exponential regime for large deviations; a purely \(e^{-ct^2}\) upper tail cannot hold for a chi-square random variable at all scales.

\section*{Reproducibility statement}
No external data are used.  The accompanying script \texttt{generate\_figures.py} recreates all figures and the numerical checks in Table~\ref{tab:noncommuting}.


\end{document}